\documentclass[review]{elsarticle}

\usepackage{lineno,hyperref}
\modulolinenumbers[5]

\journal{Journal of \LaTeX\ Templates}

\usepackage{hyperref}
\usepackage[super,sort,compress]{cite}
\usepackage{subfigure}
\usepackage{color}
\usepackage{soul}
\usepackage{amsmath}
\usepackage{amssymb}
\usepackage{booktabs}
\usepackage{graphicx}   
\usepackage{bm}
\usepackage{color}
\usepackage{cancel}
\usepackage{slashed}
\usepackage{subfigure}
\usepackage{soul}

\newcommand{\Ket}[1]{\left\vert #1\right\rangle}
\newcommand{\Bra}[1]{\left\langle #1\right\vert}

\newcommand{\BraKet}[2]{\left\langle#1\vert #2\right\rangle}

\newcommand{\ii}{\mathrm{i}}
\newcommand{\ee}{\mathrm{e}}

\newcommand{\delete}[1]{{\color{red} \st{#1}}}

\renewcommand{\delete}[1]{}

\renewcommand{\delete}[1]{{\color{blue} \st{#1}}}

\begin{document}

\title{Hilbert space partitioning for non-Hermitian Hamiltonians: from off-resonance to Zeno subspaces}

\author{Benedetto Militello}
\author{Anna Napoli}
\address{Universit\`a degli Studi di Palermo, Dipartimento di Fisica e Chimica - Emilio Segr\`e, Via Archirafi 36, I-90123 Palermo, Italy}
\address{I.N.F.N. Sezione di Catania, Via Santa Sofia 64, I-95123 Catania, Italy}

\begin{abstract}

Effective non-Hermitian Hamiltonians describing decaying systems are derived and analyzed in connection with the occurrence of possible Hilbert space partitioning, resulting in a confinement of the dynamics. In some cases, this fact can be interpreted properly as Zeno effect or Zeno dynamics, according to the dimension of the subspace one focuses on; in some other cases, the interpretation is more complicated and traceable back to a mix of Zeno phenomena and lack of resonance. Depending on the complex phases of the diagonal terms of the Hamiltonian, the system reacts in different ways, requiring larger moduli for the dynamical confinement to occur when the complex phase is close to $\pi/2$. 
 
\end{abstract}

\maketitle

\section{Introduction}\label{sec:introduction}

The quantum Zeno effect (QZE), in its original formulation, is the inhibition of the natural time evolution of a physical system due to repeated measurements~\cite{ref:MishraSudarshan}. In fact, the wave function collapse and the quadratic behavior of the survival probability (the probability to find the system in its initial state) of a quantum system at short time, both avoid any dynamical evolution of a frequently observed system. This paradigmatic effect underlines the active role of measurements in quantum mechanics and has been experimentally demonstrated in connection with Rabi oscillations in trapped ions~\cite{ref:Itano1990} and tunnel effects in confined atoms~\cite{ref:Fischer1997}. Experimental proofs of the quantum Zeno effect have been provided in the context of Bose-Einstein condensates~\cite{ref:Schafer2014,ref:Streed2006}. The original formulation has been gradually extended including ways to act on a quantum system different from proper measurements. Indeed, for example, a decaying quantum state can be interpreted as a state which is continuously observed by an environment: if a photon is observed that has been emitted in the decay process of a quantum state, then we can say that the system was in the decaying state~\cite{ref:Presilla1996,ref:Home1997,ref:Schulman1998,ref:PascazioFacchi2001,ref:PascazioFacchi2008,ref:PascazioFacchi2010}. In fact, a strong decay is proven to play the same role of frequent measurements, hence hindering the time evolution~\cite{ref:Panov1999}. Of course, in such a situation the inhibition can also be interpreted as a consequence of a dynamical decoupling, which has been predicted in several physical contexts, from the STIRAP manipulation~\cite{ref:ScalaPRA2010,ref:ScalaOpts2011,ref:MilitelloPScr2011} to the quantum biological processes~\cite{ref:Caruso} to spin-chain systems~\cite{ref:Popkov2018}. 
Since a decay is the consequence of an interaction between one level and a continuum of levels, the subsequent most natural extension concerns the case where a coupling induces a Hilbert space partitioning responsible for making ineffective some other interactions~\cite{ref:Militello2001,ref:PascazioFacchi2002PRL}. It is interesting to note that the same occurrence can be found in completely classical systems~\cite{ref:Peres}.
When the external agents (frequent measurements, strong decays or intense interactions) isolate a degenerate subspace one has the quantum Zeno dynamics. The subspace where the system is repeatedly projected undergoes a dynamics which does not take into account the interactions connecting this subspace to others. The Zeno effect is then a special case of Zeno dynamics with a trivial dynamics.

It is worth mentioning that when the measurements are frequent but not frequent enough an acceleration of the dynamics of the system can occur, instead of an inhibition, leading to the anti-Zeno effect (AZE). The boundary between QZE and AZE has been extensively studied and in the case of a system subjected to an interaction with an environment, the AZE-QZE threshold is traceable back to the spectral properties of the environment~\cite{ref:Kofman2000,ref:Kofman2001} and is influenced by the temperature~\cite{ref:Maniscalco2006} as well as by the bath statistics~\cite{ref:Rao2011}. Temperature can have an important role in the occurrence of Zeno phenomena. In fact, a certain influence of the detector's temperature on the Zeno effect has been predicted~\cite{ref:Ruse2002}, as well as a role of temperature in continuous measurement QZE associated to the system-environment interaction~\cite{ref:Militello2011,ref:Militello2012}. Moreover, thermodynamic processes can be influenced by quantum Zeno phenomena~\cite{ref:Erez2008}. It is worth mentioning that a behavior similar to that classified as Zeno dynamics can be obtained any time a Hilbert space partitioning occurs, i.e. every time some interaction terms are rendered ineffective for some reasons. A very specific example is given by the presence of a  large energy gap, which can bring out of resonance any coupling. In the general case, both the lack of resonance and the Zeno occurrences can contribute to the inhibition of the dynamics, leading to a hybrid situation. The mathematical counterpart of such a general situation is the introduction of non-Hermitian Hamiltonians with complex diagonal entries, taking into account of both the presence of an energy gap (through their real parts) and the presence of decays (through the imaginary parts). By the way, the wider phenomenology of a Hilbert space partitioning and consequent inhibition of the dynamics due to either or both causes will be addressed in the following as \lq extended Zeno dynamics\rq\, (EZD), including as special cases the pure Zeno effect and the pure lack of resonance, the latter emerging from a Hermitian Hamiltonian.
The link between quantum Zeno effect and non-Hermitian Hamiltonians is deeper than expected. Indeed, on the one hand, the QZE induced by strong decays has been studied through non-Hermitian Hamiltonian models~\cite{ref:Home1997,ref:Schulman1998,ref:Muga2008,ref:Echa2008}, while, on the other hand, the effects of repeated measurements have been proven to be describable via suitable non-Hermitian effective Hamiltonians~\cite{ref:Koz2016}. The relevance of having or not ${\cal PT}$-symmetry on the occurrence of the Zeno and anti-Zeno effect has been investigated~\cite{ref:Naikoo2019}.
In spite of these connections and of the growing interest in non-Hermitian Hamiltonians~\cite{ref:Bender1998,ref:Rudner2009,ref:Feng2011,ref:Regensburger2012,ref:Fyod2012,ref:Gros2014,ref:Ashida2017,ref:Nakagawa2018,ref:Kawabata2019}, there is not a systematic study non model-dependent of the quantum Zeno effect in the presence of a non-Hermitian Hamiltonian.

In this paper we analyze a physical scenario where a set of possibly decaying levels are coupled to a set of non-decaying ones. When the gap between the two subspaces (in terms of complex diagonal entries) is very large, we get an EZD, irrespectively of the phases of the diagonal entries of the effective non-Hermitian Hamiltonian describing the system.
On the contrary, when the gap is moderately large, the system becomes very sensitive to the phase of the diagonal entries and the occurrence of an EZD  requires higher values of the gap when the phases are close to $\pi/2$. The introduction of proper indicators allows to bring to light extended Zeno dynamics even in some regimes where is seemingly absent. Insensitivity of some levels to the interaction with other levels is also possible, in connection with special initial conditions, when the requirements for an interaction-free evolution (IFE) are fulfilled~\cite{ref:Chrusc2014,ref:Chrusc2015,ref:Chrusc2016}.
In the next section we introduce the Hamiltonian model for a system with a group of levels which undergo decays toward levels external to the subspaces we are focusing on. We explicitly prove that under suitable hypotheses the system can be properly described by an effective non-Hermitian Hamiltonian equivalent to the relevant master equation. In the same section, we also apply the perturbation theory to the case where large gaps are present in the complex spectrum of the non-Hermitian Hamiltonian. In sec.~\ref{sec:Zeno} we analyze the EZD induced by large gaps between the diagonal entries of the Hamiltonian. We first report on some analytical arguments, then, in section \ref{sec:Three-State}, we analyze some numerical results obtained for the specific case of a three-state Hamiltonian with a decaying level. Finally, in sec.~\ref{sec:Conclusions} we give some conclusive remarks.

\section{Non-Hermitian Hamiltonians}\label{sec:non-Hermitian}

The appearance of non-Hermiticity in Hamiltonian operators is always traceable back to the derivation of an effective Hamiltonian which takes into account the interaction with external degrees of freedom not explicitly included in the description of the reduced system. In the next subsection we consider the effective Hamiltonian description of a system with decaying states, leading to complex diagonal terms, the real parts being the proper energies of the relevant levels, whereas the imaginary parts are the decay rates~\cite{ref:Severini2004,ref:ScalaPRA2010,ref:Militello2016}. In the subsequent subsection we apply the perturbation theory to our non-Hermitian Hamiltonian in a special regime, i.e., when two well separated bands associated to the bare \lq complex energies\rq\, (i.e., the diagonal terms) can be identified.

\subsection{Non-Hermitian Hamiltonian for decaying systems}

\begin{figure}[h]
\centering
\includegraphics[width=0.60\textwidth, angle=0]{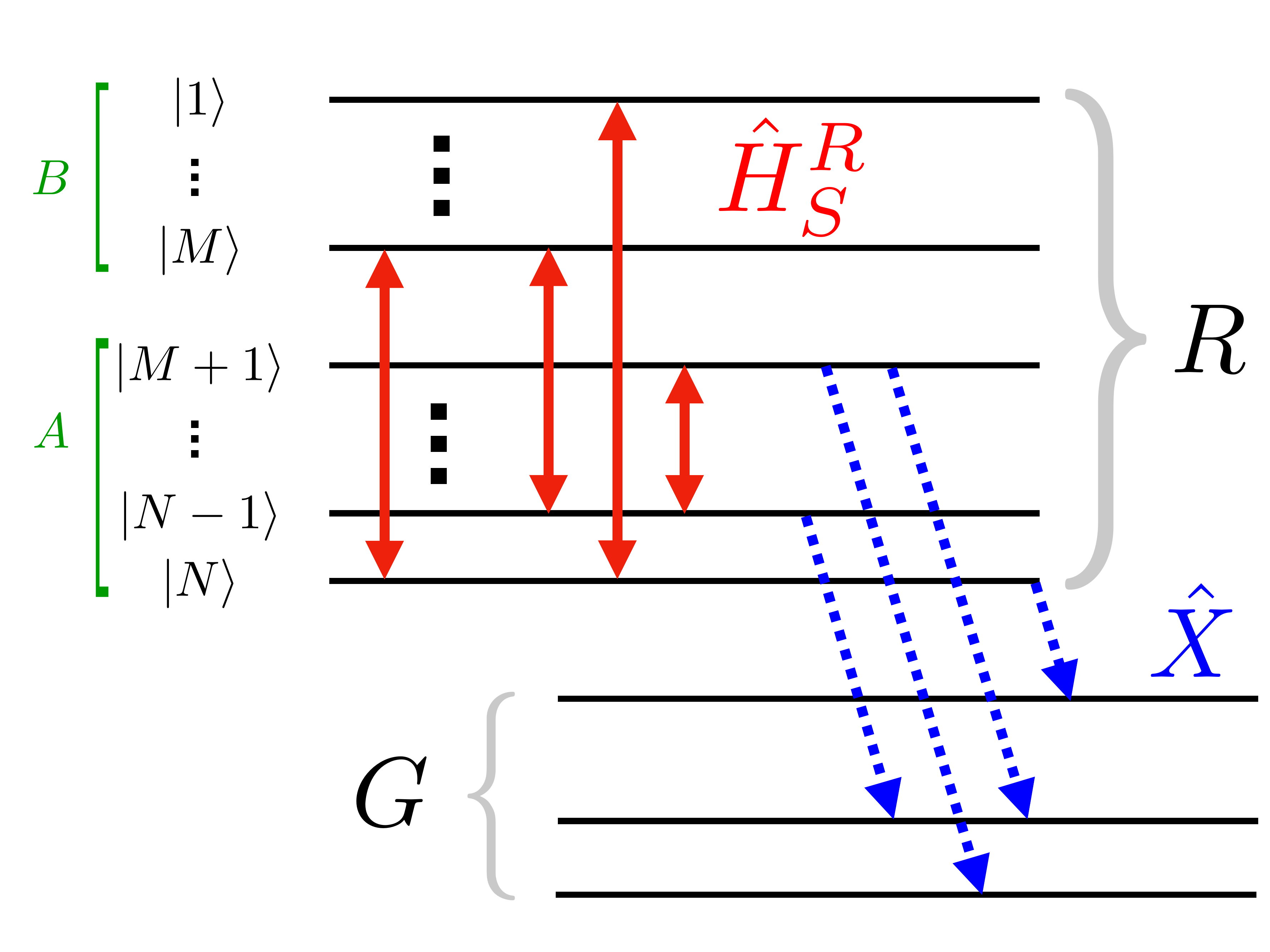}
\caption{(Color online) Scheme of the coherent and incoherent couplings. The states of the subspace $R$ are coherently coupled to each other  by $\hat{H}_S$ (red solid two-side arrows) and part of them, from $M+1$ to $N$, are coupled through the operator $\hat{X}$ of the system-environment coupling term to some states of the subspace $G$ (blue dotted one-side arrows). States from $1$ to $M$ constitute the subspace $A$ (states undergoing decays), while the states from $M+1$ to $N$ belong to the subspace $B$ (states not undergoing decays).}\label{fig:CouplingScheme}
\end{figure}

A system with some states undergoing decay processes toward some lower states due to the interaction with a zero-temperature environment, can be described through an effective non-Hermitian Hamiltonian, provided we focus on a subspace not involving the states receiving population from the decaying ones~\cite{ref:Severini2004,ref:ScalaPRA2010,ref:Militello2016}. 
In fact, we consider a system whose Hilbert space can be decomposed in three subspaces, $A$ and $B$, together forming $R=A\oplus B$, and the subspace of the lowest band $G$. Only the subspace $A$ is coupled to $G$ through the environment. Moreover, states of $A$ and $B$ are coherently coupled to each other, but not coherently coupled to $G$. Fig.~\ref{fig:CouplingScheme} illustrates this situation.
More in detail, let $\hat{H}_S$ be the Hamiltonian governing the dynamics of $R\oplus G$, while $\hat{\Pi}_R$ and $\hat{\Pi}_G$ are the projectors onto the two subspaces.
Now, assume a system-environment interaction term $\hat{H}_{SE}=\lambda \hat{X} \otimes \hat{E}$, with $\hat{X} = \hat{\Pi}_{R} \hat{X} \hat{\Pi}_{G} + \hat{\Pi}_{G} \hat{X} \hat{\Pi}_{R}$, so that incoherent transitions within $R$ are excluded: $ \hat{\Pi}_{R} \hat{X} \hat{\Pi}_{R}=0$. According to the general theory of open quantum systems~\cite{ref:Gardiner,ref:Petru}, the relevant zero-temperature Markovian master equation can be written as:
\begin{equation}\label{eq:NStateSystemME}
\dot\rho = -\ii [\hat{H}_S, \rho] + \sum_{ij} \gamma_{ij} \left( \hat{X}_{ij} \rho \hat{X}_{ij}^\dag - \frac{1}{2} \{ \hat{X}_{ij}^\dag \hat{X}_{ij}, \rho\} \right) \,,
\end{equation}
where $\hat{X}_{ij}$ are suitable jump operators connecting states of $R$ with states of $G$, $\gamma_{ij}$ being the relevant decay rates. The Lamb-shifts have been neglected. Because of the zero-temperature assumption, the structure of $\hat{X}$ and the fact that the energies of $G$ are lower than those of $R$, only the terms with $\hat{X}_{ij} = \hat{\Pi}_{G} \hat{X}_{ij} \hat{\Pi}_{R}$ have non vanishing $\gamma_{ij}$, while the terms with $\hat{X}_{ij} = \hat{\Pi}_{R} \hat{X}_{ij} \hat{\Pi}_{G}$ are absent.

Now, if we assume also that $\hat{H}_S$ does not couple the subspaces $R$ and $G$ (i.e., assume $\hat{H}_S = \hat{\Pi}_{R} \hat{H}_S \hat{\Pi}_{R} + \hat{\Pi}_{G} \hat{H}_S \hat{\Pi}_{G} $), a closed equation for the density operator restricted to the subspace $R$ can be straightforwardly obtained. Indeed, introducing $\rho^{R} \equiv \hat{\Pi}_{R} \rho \hat{\Pi}_{R}$, one gets,
\begin{equation}\label{eq:NStateSystem}
\dot\rho^{R} = -\ii [\hat{H}_S^{R}, \rho^{R}] - \sum_{ij} \gamma_{ij} \frac{1}{2} \{ \hat{X}_{ij}^\dag \hat{X}_{ij}, \rho^{R}\} \,,
\end{equation}
with $\hat{H}_S^{R} \equiv \hat{\Pi}_{R} \hat{H}_S \hat{\Pi}_{R}$ and where we have used both $\hat{\Pi}_{R} \hat{X}_{ij} \hat{\Pi}_{R} = 0$\, $\forall i, j$ and $\hat{X}_{ij}^\dag \hat{X}_{ij} = \hat{\Pi}_{R} \hat{X}_{ij}^\dag \hat{X}_{ij} \hat{\Pi}_{R}$.

This equation can be put in the form of a pseudo-Liouville-von Neumann equation, 
\begin{equation}\label{eq:PseudoLiouville}
\dot\rho^{R} = -\ii (H \rho^{R} - \rho^{R} H^\dag)\,, 
\end{equation}
with the non-Hermitian Hamiltonian
\begin{equation}\label{eq:NStateSystemRedME}
H = \hat{H}_S^{R} - \ii \sum_{ij} \frac{\gamma_{ij}}{2} \hat{X}_{ij}^\dag \hat{X}_{ij} \,.
\end{equation}
The restriction to $R$ of the state at time $t$ is easily evaluated as
\begin{equation}\label{eq:Evolution}
 \rho^{R}(t) = \ee^{-\ii H t} \rho^{R}(0) \ee^{\ii H^\dag t}\,.
\end{equation}
This solution of \eqref{eq:PseudoLiouville} is valid for every initial state $\rho^R(0)$, whether pure or not. On the other hand, if the initial state is pure $\rho^{R}(0)=\Ket{\psi(0)}\Bra{\psi(0)}$, this expression turns out to be equivalent to writing $\Ket{\psi(t)}= \ee^{-\ii H t} \Ket{\psi(0)}$, solution of the pseudo-Schr\"odinger equation $\ii\partial_t\Ket{\psi}=H\Ket{\psi}$ involving the non-Hermitian Hamiltonian $H$.

\begin{figure}[h]
\centering
\includegraphics[width=0.60\textwidth, angle=0]{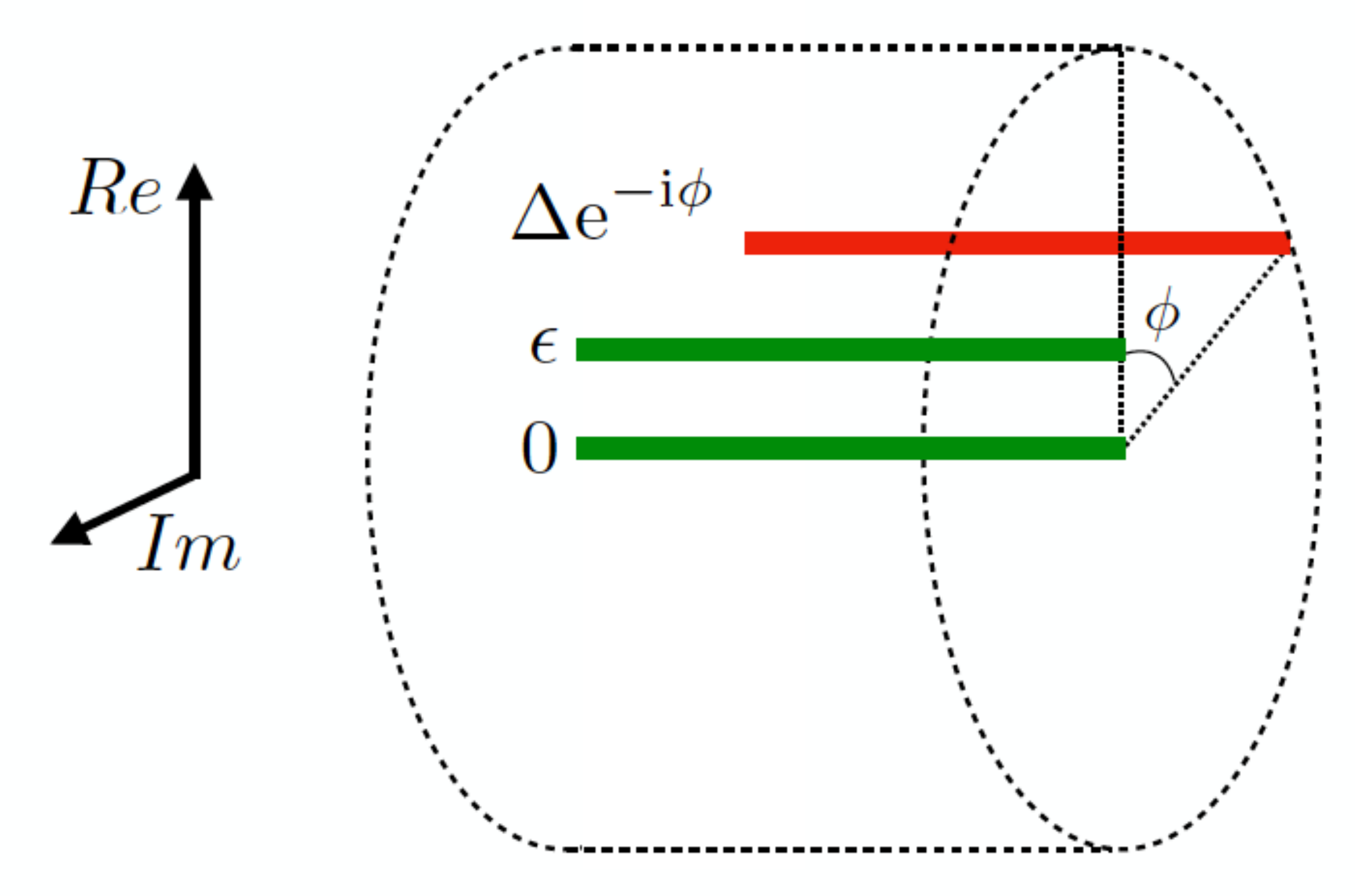}
\caption{(Color online) Levels of a three-state system with complex energies $0$, $\epsilon$ (real numbers) and $\Delta \ee^{-\ii\phi}$ (complex value). The parameter $\phi$ must lie in the interval $[0,\pi]$ to prevent positive imaginary parts of $\Delta \ee^{-\ii\phi}$.}\label{fig:LevelScheme}
\end{figure}

Let us now introduce the eigenstates of the operator $\sum_{ij} \frac{\gamma_{ij}}{2} \hat{X}_{ij}^\dag \hat{X}_{ij}$, denote them as $\Ket{k}$, and rewrite the Hamiltonian in the following form:
\begin{equation}\label{eq:NStateSystemH}
H = \sum_k \Delta_k \ee^{-\ii\phi_k} \Ket{k}\Bra{k}  + \sum_{j\not=k} h_{jk} \Ket{j}\Bra{k}\,.
\end{equation}
Here the diagonal entries, either expressed in terms of moduli and phases as $\Delta_k \ee^{-\ii\phi_k}$ or written in terms of real and imaginary parts as $\epsilon_k -\ii \Gamma_k$, contain both information about the energies of the levels (real parts) and the relevant decay rates (imaginary parts). We can address them as \lq complex energies.\rq\, 
The off-diagonal terms $h_{ij}$ represent the coupling strengths between different eigenstates.
In Fig.~\ref{fig:LevelScheme} it is shown the paradigmatic case where a three-state system is characterized by two energy levels ($0$ and $\epsilon$) which do not decay and a third level which decays, then having a complex energy $\Delta\ee^{-\ii\phi}$. Two non-decaying states and a decaying one are the minimal requirement to have Zeno dynamics (instead of a simple Zeno effect). In sec.~\ref{sec:Three-State} we focus on this specific situation.

Since we have assumed that only the states of the subspace $A$ are coupled to states of $G$, the effective system Hamiltonian can be rewritten as $H=H_0+H_I$, with
\begin{eqnarray}\label{eq:DefH0HI}
H_0 = 
\left(\begin{array}{cc}
\mathbf{A} & \mathbf{0} \\
\mathbf{0} & \mathbf{B}
\end{array}\right)\,, \qquad 
H_I = 
\left(\begin{array}{cc}
\mathbf{0} & \mathbf{C} \\
\mathbf{C}^\dag & \mathbf{0}
\end{array}\right)\,.
\end{eqnarray}
where $\mathbf{B}$ is an Hermitian operator and $\mathbf{A}$ is a non-Hermitian matrix that can be assumed to be in a diagonal form. 
We emphasize that $\mathbf{A}$ and $\mathbf{B}$ are the restrictions of the Hamiltonian $H$ to the subspaces $A$ and $B$, respectively, while $\mathbf{C}$ contains the coupling terms between eigenstates belonging to the two subspaces.
Since $\mathbf{B}$ is Hermitian, it can always be diagonalized. Therefore, we will assume that the diagonalization has already been carried on.

It is the case to comment on the fact that introducing the Hamiltonian $H$ instead of keeping the whole master equation in \eqref{eq:NStateSystemME} is not an approximation, when the system starts in the subspace $R$. In fact, it is a simplification for treating the dynamics in the subspace $R$ by dealing with a square matrix of order $\mathrm{tr}\hat{\Pi}_R$ (the non-Hermitian Hamiltonian) acting on a Hilbert space instead of working with a square matrix of order $(\mathrm{tr}\hat{\Pi}_R + \mathrm{tr}\hat{\Pi}_G)^2$ (the representation of the superoperator in \eqref{eq:NStateSystemME}) acting on the relevant Liouville space (see \ref{sec:appME} for details).

\subsection{Perturbation treatment}\label{sec:perturbation}

Let us consider the Hamiltonian $H=H_0+H_I$ with $H_0$ and $H_I$ given by \eqref{eq:DefH0HI}, under the assumption that the coupling between the two subspaces $A$ and $B$ is weak, meaning that the coupling strengths between the subspaces are small when compared to the two-band energy gap:
\begin{eqnarray}
|a_{nn} - b_{mm}| \gg |c_{ij}| \,, \qquad \forall i, j, m, n\,,
\end{eqnarray}
where $a_{nm}$, $b_{nm}$ and $c_{ij}$ are the entries of $\mathbf{A}$, $\mathbf{B}$ and $\mathbf{C}$, respectively.
Introducing $\delta = \min_{nm}\{|a_{nn} - b_{mm}|\}$ and $c=\max_{ij}|c_{ij}|$, we can rewrite this condition as
\begin{eqnarray}\label{eq:separation}
c/\delta \ll 1.
\end{eqnarray}

Under such hypothesis, an approximated diagonalization of $H$ can be carried on quite easily through the perturbation theory. Since we are dealing with a non-Hermitian Hamiltonian, some delicate points have to be taken into account (details of this treatment are reported in the \ref{app:perturbation}). The first order-corrected eigenvalues and eigenvectors turn out to be (all the $m$ indexes span the $A$ subspace, then ranging from $1$ to $M$, while all the $n$ indexes span the $B$ subspace, then ranging from $M+1$ to $N$):
\begin{subequations}
\begin{eqnarray}
\alpha_m &=& \Delta_m \ee^{-\ii\phi_m} \,, \\ 
\Ket{\alpha^R_m} &=& \Ket{m} + \sum_n \frac{c_{mn}}{\Delta_m \ee^{-\ii\phi_m}-E_n} \Ket{n} \, ,  \\
\Bra{\alpha^L_m} &=& \Bra{m} + \sum_n \frac{c_{nm}}{\Delta_m \ee^{-\ii\phi_m}-E_n} \Bra{n} \, ,  \\
\beta_n &=& E_n \,, \\
\Ket{\beta^R_n} &=& \Ket{n} + \sum_m \frac{c_{nm}}{E_n - \Delta_m \ee^{-\ii\phi_m}} \Ket{m} \, ,  \\
\Bra{\beta^L_n} &=& \Ket{n} + \sum_m \frac{c_{mn}}{E_n - \Delta_m \ee^{-\ii\phi_m}} \Ket{m} \, . 
\end{eqnarray}
\end{subequations}

It is important to note that $\Bra{\alpha^L_m}\not=(\Ket{\alpha_m^R})^\dag$ and $\Bra{\beta^L_n}\not=(\Ket{\beta_n^R})^\dag$. Indeed, though $c_{mn}=c_{nm}^*$, the denominators of the first-order correction are the same complex number for  $\Bra{\alpha^L_m}$ and $\Ket{\alpha_m^R}$ (i.e., $\Delta_m \ee^{-\ii\phi_m}-E_n$) not the complex conjugate to each other.

Concerning the second order correction, we focus on the eigenvalues (for the eigenstates see \ref{app:perturbation}):
\begin{subequations}
\begin{eqnarray}
\alpha_m &=& \Delta_m \ee^{-\ii\phi_m} + \sum_{n} \frac{|c_{nm}|^2}{\Delta_m \ee^{-\ii\phi_m} - E_n} \,, \\
\beta_n &=& E_n +  \sum_{m} \frac{|c_{nm}|^2}{E_n - \Delta_m \ee^{-\ii\phi_m}} \,.
\end{eqnarray}
\end{subequations}

It is interesting to note that the corrections to the real energies in the $B$ subspace are complex numbers, meaning that decay processes occur also in the subspace which is subjected to a unitary dynamics in the unperturbed case.
In particular, 
\begin{eqnarray}\label{sec:effdecay}
\mathrm{Im}{\beta_n} = - \sum_{m} \frac{|c_{nm}|^2 \Delta_m }{|E_n - \Delta_m \ee^{-\ii\phi_m}|^2} \sin{\phi_m} \,
\end{eqnarray}
is the effective decay rate associated to the state $\Ket{\beta^R_n}$ obtained by correcting the state $\Ket{n}$ of the subspace $B$.
This quantity is expected to be higher when $\phi_m\approx \pi/2$ $\forall m$, and smaller when $\phi_m\approx 0, \pi$ $\forall m$. On this basis, we can expect a role of the phases $\phi_m$ on the appearance of an extended Zeno dynamics when the moduli $\Delta_m$ are moderately large.

\section{Extended Zeno dynamics}\label{sec:Zeno}

In this section we investigate the occurrence of an extended Zeno dynamics in a system governed by an Hamiltonian as in \eqref{eq:NStateSystemH} which in general is non-Hermitian, and under the assumption of sec.~\ref{sec:perturbation} that there is a large gap between the two subspaces $A$ and $B$. It is known that a decay can play the role of continuous measurement on a quantum system, and when the relevant decay rate is large enough (which is the continuous counterpart of getting a larger number of measurements in a given time interval) a partitioning of the Hilbert space can produce either a pure Zeno effect (freezing the system in its initial condition) or a pure Zeno dynamics. Therefore, when we have $M$ levels which have $\Delta_k \ee^{-\ii\phi_k} = -\ii\Gamma_k$ very large with respect to all the other parameters, the states in the subspace $B$ evolve as if no interaction between the first subspace and the second one were present.
Similarly, when there are very large real diagonal elements (a set of $M$ states with very large $\Delta_k$'s and $\phi_k=0,\pi$) we have that the dynamics of the relevant states is well separated from the dynamics of the remaining $N-M$ ones, and again the dynamics of this second subspace is the one obtained in the absence of any interaction with the first subspace.

In the following we investigate the more general situation where a set of $\Delta_k$'s ($k=1,..., M$) are very large while the phases can assume any value. In particular, we want to investigate whether the extended Zeno dynamics occurs irrespectively of $\phi_k$'s.

{ \it Perturbed vs unperturbed dynamics } --- Assuming that the system starts in a certain state $\Ket{\psi(0)}$, we evaluate the two evolutions given by the equations $\ii\partial_t \Ket{\psi_0(t)} = H_0 \Ket{\psi_0(t)}$ (unperturbed evolution) and $\ii\partial_t \Ket{\psi(t)} = (H_0 + H_I) \Ket{\psi(t)}$ (perturbed evolution). The relevant solutions are:
\begin{eqnarray}
\label{eq:Psi_0_t}
\Ket{\psi_0(t)} &=& \sum_n \BraKet{n}{\psi(0)} \ee^{-\ii E_n t } \Ket{n} \,,\\
\label{eq:Psi_t}
\Ket{\psi(t)} &=&  \sum_m \BraKet{\alpha_m^L}{\psi(0)} \ee^{-\ii \alpha_m t } \Ket{a_m^R} 
+ \sum_n \BraKet{\beta_n^L}{\psi(0)} \ee^{-\ii \beta_n t } \Ket{\beta_n^R} \,.
\end{eqnarray}

When condition in \eqref{eq:separation} is fulfilled then the perturbation treatment is allowed and, moreover, the smaller $c/\delta$, the smaller the corrections to the eigenvalues and eigenvectors, the closer the evolutions induced by $H=H_0+H_I$ and $H_0$ are. When $c/\delta$ is small though not extremely small, deviations between the two dynamics can be observed.
In fact, for extremely small $c/\delta$ ($\rightarrow 0$) the perturbed and unperturbed eigenvalues and eigenvectors coincide. For small but not extremely small $c/\delta$ we can consider good the first-order approximation, which leaves unchanged the eigenvalues and slightly changes the eigenvectors. The evolution is then characterized by the same frequencies and phase factors characterizing the unperturbed case, but corresponding to slightly different states. For moderately small values of $c/\delta$ the second-order correction is more appropriate, leading to corrections of the eigenvalues, which in general become complex also in the $B$ subspace, as given by \eqref{sec:effdecay}. In this case, the unitary dynamics in the $B$ subspace is replaced by a non-unitary one and a general loss of probability in this subspace is predicted.
It is worth emphasizing that if the corrected eigenvalues were real numbers, the discrepancy between the unperturbed and the perturbed dynamics would be $o(c/\delta)$ at any time. The presence of imaginary parts in the eigenvalues makes the gap between the two evolutions increase with time, due to the presence of negative exponentials. 
This is a crucial point since it introduces an extremely different behavior depending on the phases of the diagonal terms. In fact, the Hilbert space partitioning occurring for \lq real energies\rq\, ($\phi_m\approx 0, \pi$) produces a confinement of the dynamics in the subspace $B$ without dissipation, that is a long-standing EZD, while the decays present when \lq complex energies\rq\, are considered produce a time-increasing deviation between the complete dynamics and the unperturbed one. Indeed, while in the former case $\mathrm{Im}\beta_n = 0$, in the latter case $\mathrm{Im}\beta_n \not= 0$, which implies a loss of probability (see \eqref{eq:Psi_t}) and then an inevitably growing discrepancy between $\Ket{\psi_0(t)}$ and $\Ket{\psi(t)}$ as time goes on. Nevertheless, normalization of the wave vector can give a state very close the one obtained through the unperturbed evolution. This effect is more significant if the imaginary parts of the corrected energies turn out to be all equal to each other, $\mathrm{Im}\beta_n = \mathrm{Im}\beta_{n'}\, \forall n, n'$, because in such a case the exponential factors do not introduce any distortion of the dynamics, but are responsible only for a global loss of probability expressible through a global exponential factor $\ee^{-\Gamma t}$ (with $\Gamma=\mathrm{Im}\beta_n$) to the wave vector. 

Finally, note that the quantities $\mathrm{Im}\beta_n$ can be made smaller and smaller by increasing the values of the $\Delta_m$'s. This means that in order to have a long-standing EZD higher values of the moduli of the diagonal matrix elements are required when the energies are complex (especially if they are purely imaginary) than when they are real.

{ \it Indicators for EZD} --- In order to better analyze the appearance of an extended Zeno dynamics, we introduce suitable fidelities. 
In particular we calculate the following quantity:
\begin{equation}\label{eq:Fid}
{\cal F}(T) = \min_{t\in[0,T]} \frac{ |\Bra{\psi_0(t)}\Pi_B \Ket{\psi(t)}|^2 } {\sqrt{ | \Bra{\psi_0(0)}\Pi_B\Ket{\psi_0(0)} | \cdot |\Bra{\psi(0)}\Pi_B \Ket{\psi(0)} | }}\,,
\end{equation}
which gives us the minimum overlap between the unperturbed and the perturbed dynamics in the subspace $B$ in a time interval $[0,T]$. An extended Zeno phenomenon (whether a freezing in the initial state or a dynamics confined to a subspace) occurs in the time interval when such a quantity approaches unity.

As previously pointed out, since we are in the presence of dissipation, a loss of global probability is generally expected, thus having $\Bra{\psi_0(t)} \hat{\Pi}_B \Ket{\psi_0(t)} \le \Bra{\psi(0)} \hat{\Pi}_B \Ket{\psi(0)}$ and $\Bra{\psi(t)} \hat{\Pi}_B \Ket{\psi(t)} \le \Bra{\psi(0)} \hat{\Pi}_B \Ket{\psi(0)}$. It can then happen that the two dynamics are very close but their scalar product is smaller than unity, giving ${\cal F} < 1$. In order to take into account this fact, we have also considered the functional 
\begin{equation}\label{eq:FidNorma}
\overline{{\cal F}}(T) = \min_{t\in[0,T]} \frac{ |\Bra{\psi_0(t)}\Pi_B \Ket{\psi(t)}|^2 } {\sqrt{ | \Bra{\psi_0(t)}\Pi_B\Ket{\psi_0(t)} | \cdot |\Bra{\psi(t)}\Pi_B \Ket{\psi(t)} | }}\,,
\end{equation}
which differs from the previous one for the normalization of the two wave functions in the subspace of interest.

According to our definition of extended Zeno dynamics, when the phases of the diagonal entries are different from $0$, $\pi/2$ and $\pi$, the confinement of the dynamics in a given subspace is due to both lack of resonance and dissipation. It can be interesting to single out the specific role of the decay rate, in order to check whether a pure Zeno effect is present. In other words, one can be interested to answer the following question: given an energy gap $\epsilon_k = \Delta_k \cos\phi_k$ and a decay rate $\Gamma_k = \Delta_k \sin\phi_k$, is the relevant dynamics more confined than in the case where the same energy gap is present but the decay is absent? A possible answer could be given by a new indicator:
\begin{equation}\label{eq:FidOnlyGap}
{\tilde{\cal F}}(T) = {\cal F}(T) - \min_{t\in[0,T]} \frac{ |\Bra{\psi_0(t)}\Pi_B \Ket{\zeta(t)}|^2 } {\sqrt{ | \Bra{\psi_0(0)}\Pi_B\Ket{\psi_0(0)} | \cdot |\Bra{\zeta(0)}\Pi_B \Ket{\zeta(0)} | }}  \,,
\end{equation}
where $\Ket{\zeta(t)}$ is the state evolving according to the following Hermitian Hamiltonian: 
\begin{equation}\label{eq:NStateSystemH-Real}
\tilde{H} = \sum_k \Delta_k \cos{\phi_k} \Ket{k}\Bra{k}  + \sum_{j\not=k} h_{jk} \Ket{j}\Bra{k}\,,
\end{equation}
obtained from $H$ by replacing the complex diagonal entries with their real parts, that is $\tilde{H}=(H + H^\dag)/2$. Of course, we require the initial condition $\Ket{\zeta(0)}=\Ket{\psi(0)}$ to be satisfied.
When ${\tilde{\cal F}}(T) > 0$ we have that the fidelity between the states $\Ket{\psi(t)}$ and $\Ket{\psi_0(t)}$ (i.e., the state evolving in the presence of both couplings and decays, and the state evolving in the absence of couplings, respectively) is higher than the fidelity between $\Ket{\zeta(t)}$ (evolving in the absence of decays but in the presence of couplings) and $\Ket{\psi_0(t)}$. Thus, the decays significantly help to confine the dynamics and neutralize the couplings, which leads to an interpretation where a proper Zeno effect is identified. On the contrary, when ${\tilde{\cal F}}(T) < 0$ we are in the presence of a more significant abandonment of the subspace $B$ because of the decays, which we could refer to as an \lq extended Anti-Zeno effect\rq.

\section{Extended Zeno dynamics in a Three-state system}\label{sec:Three-State}

We now focus on a three-state system described by the following Hamiltonian written in the basis $\{\Ket{1},\Ket{2},\Ket{3}\}$:
\begin{equation}\label{eq:ThreeStateSystem}
H = \left(
\begin{array}{ccc}
\Delta \ee^{-\ii\phi} & g_1 & g_2 \\
g_1 & \epsilon & \Omega \\
g_2 & \Omega & 0 \\
\end{array}
\right)\,.
\end{equation}
In this case the subspace $A$ consists of $\Ket{1}$ while the subspace $B$ is generated by $\Ket{2}$ and $\Ket{3}$.

\begin{figure*}[t]
\centering
\begin{tabular}{lr}
\begin{tabular}{cccc}
\subfigure[]{\includegraphics[width=0.30\textwidth, angle=0]{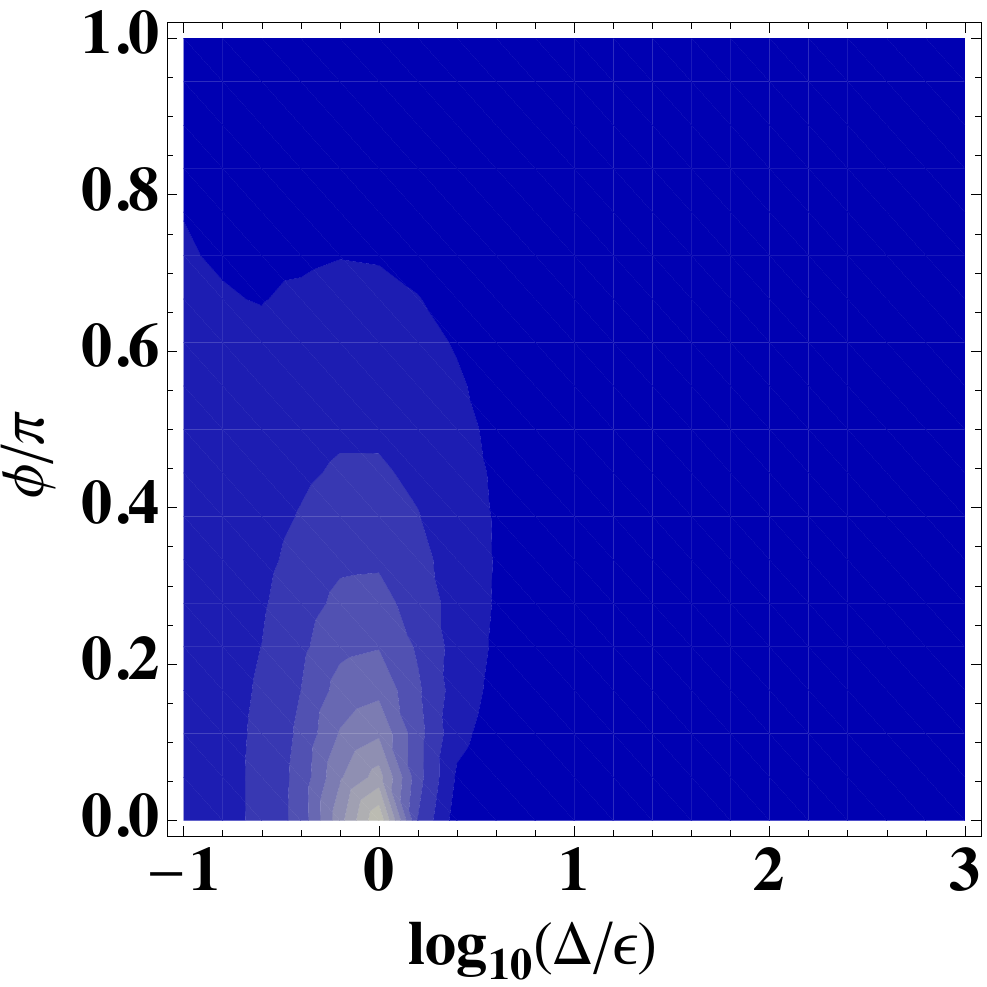}} &
\subfigure[]{\includegraphics[width=0.30\textwidth, angle=0]{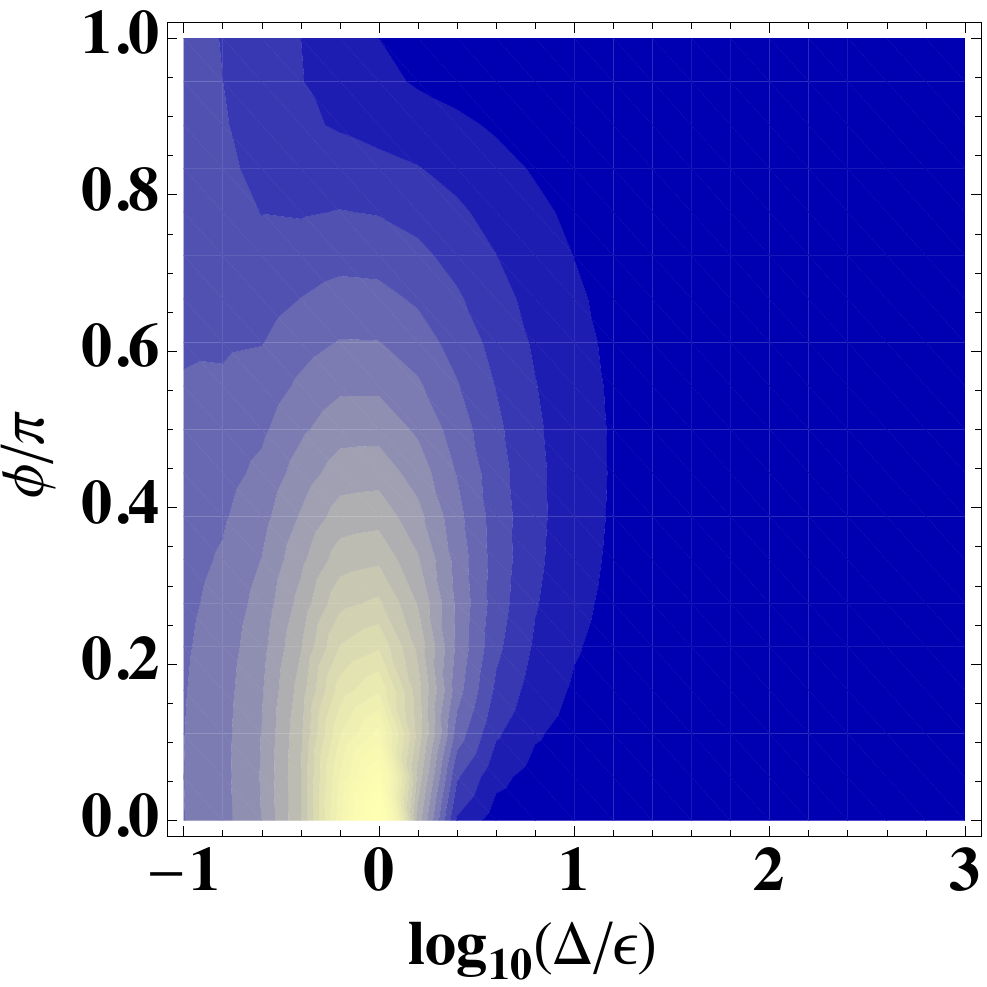}} \\
\subfigure[]{\includegraphics[width=0.30\textwidth, angle=0]{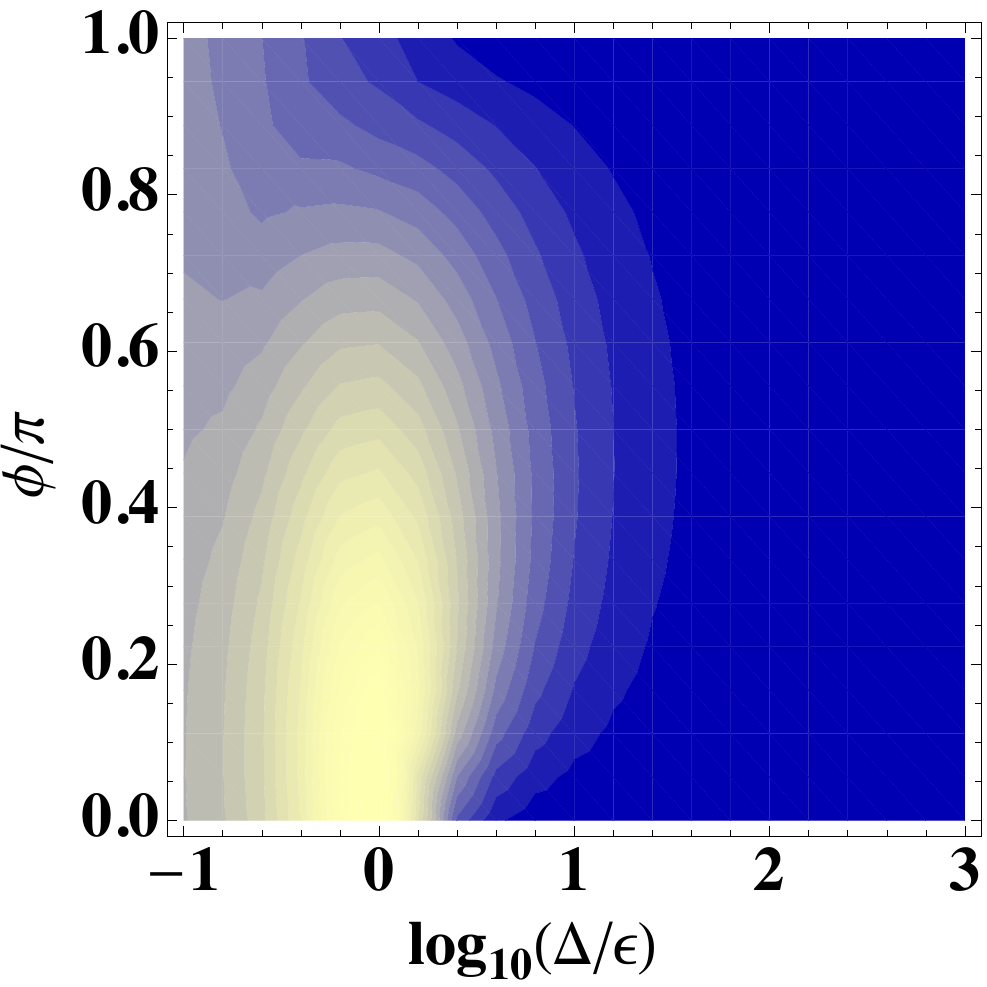}} &
\subfigure[]{\includegraphics[width=0.30\textwidth, angle=0]{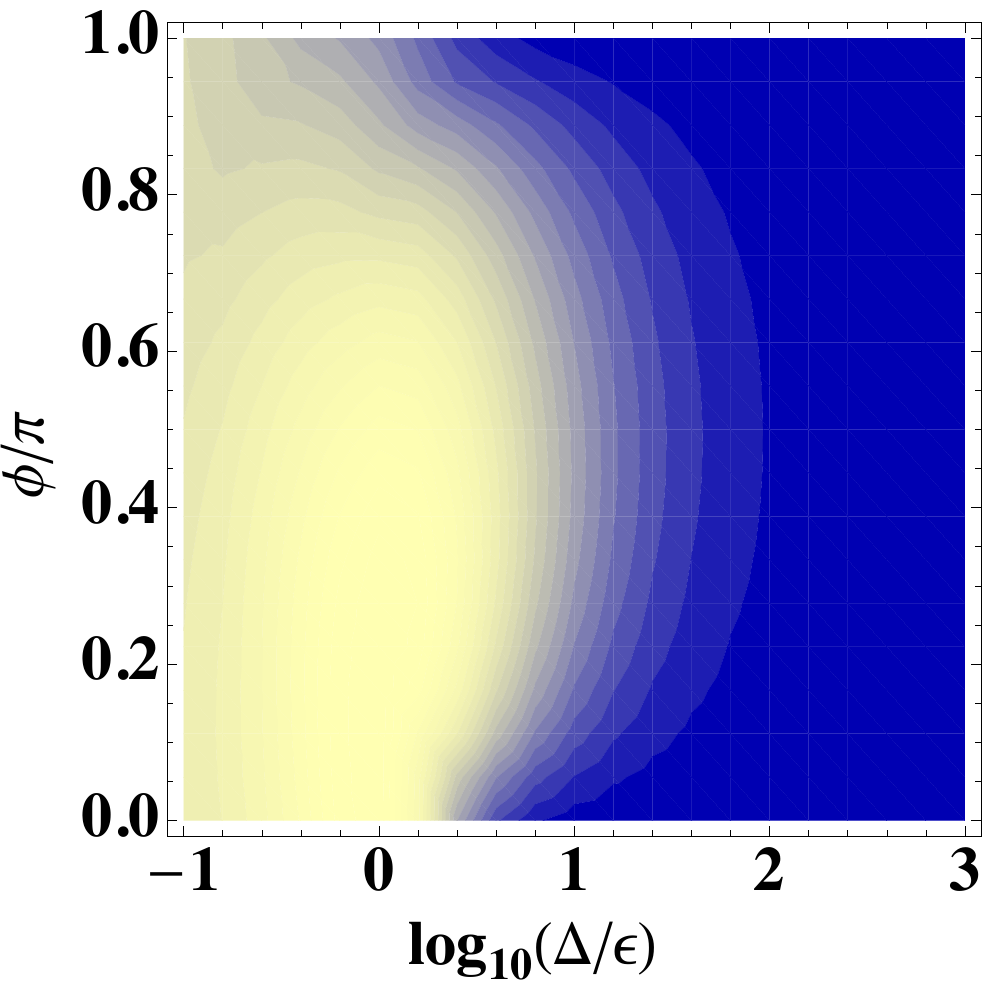}} 
\end{tabular} & 
\begin{tabular}{c}
\subfigure{\includegraphics[width=0.08\textwidth, angle=0]{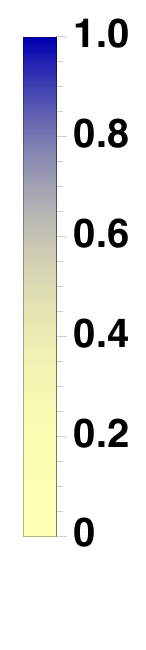}}
\end{tabular}
\end{tabular}
\caption{(Color online) 
Fidelity ${\cal F}$ as functions of $\Delta$ and $\phi$, for different values of the coupling constants: $g_1/\epsilon=g_2/\epsilon=0.1$ (a), $g_1/\epsilon=g_2/\epsilon=0.2$ (b), $g_1/\epsilon=g_2/\epsilon=0.3$ (c) and $g_1/\epsilon=g_2/\epsilon=0.5$ (d) . In all figures $\Ket{\psi(0)}=\Ket{2}$, $\Omega/\epsilon=0.1$ and $\epsilon T=2\pi$}
\label{fig:Fid_InitCond}
\end{figure*}

\begin{figure*}[t]
\centering
\begin{tabular}{lr}
\begin{tabular}{cccc}
\subfigure[]{\includegraphics[width=0.30\textwidth, angle=0]{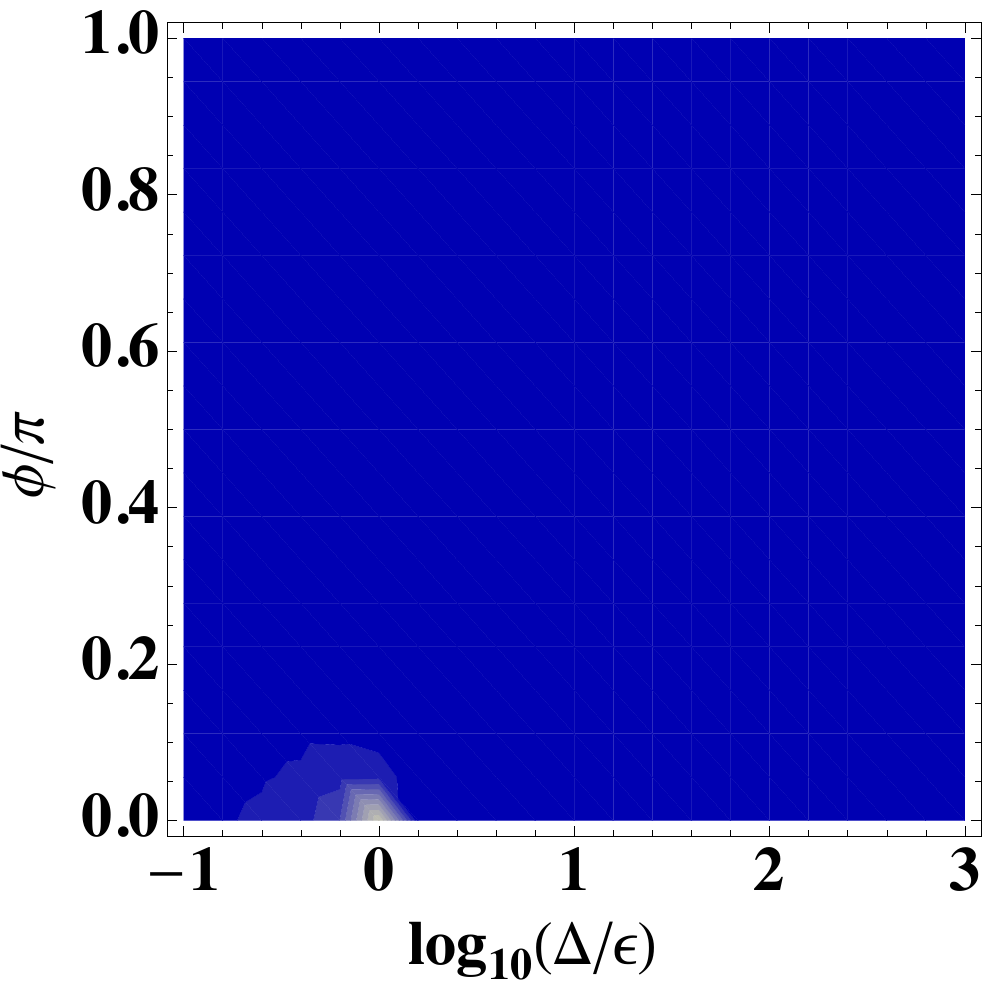}} &
\subfigure[]{\includegraphics[width=0.30\textwidth, angle=0]{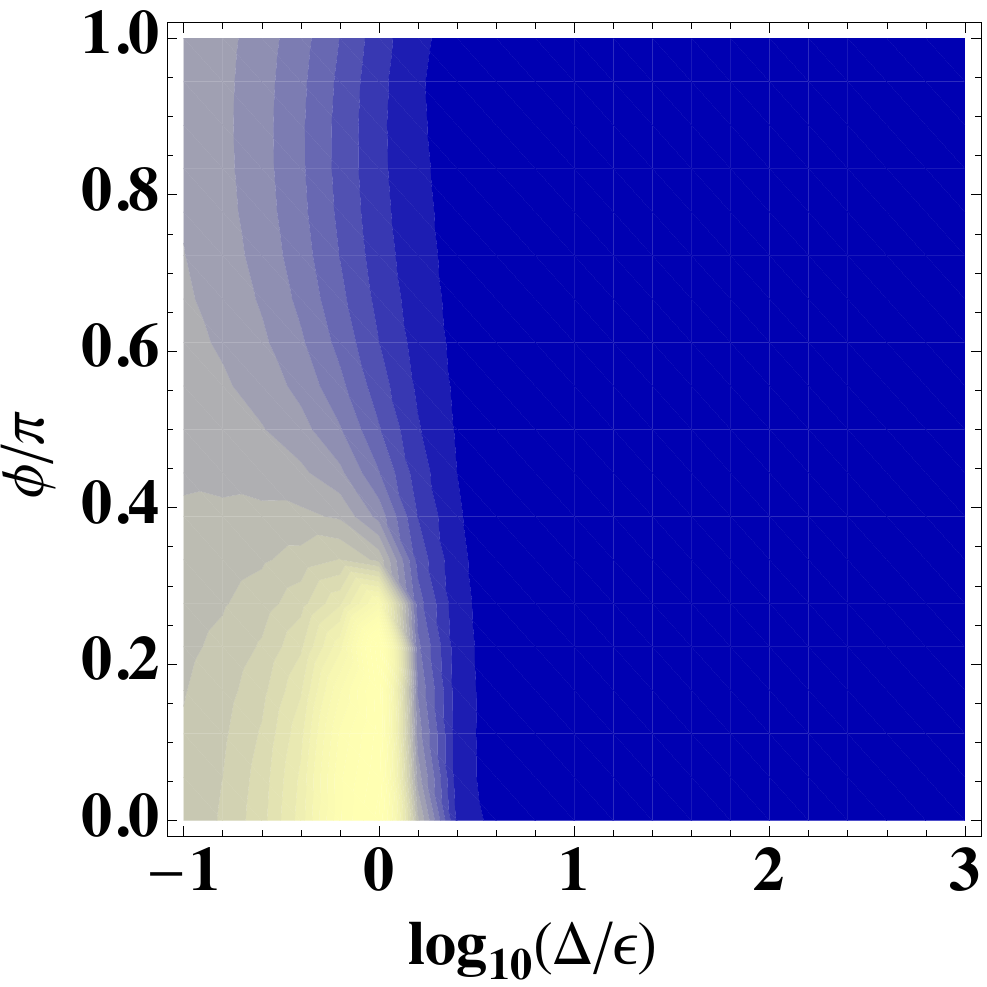}}
\end{tabular} & 
\begin{tabular}{c}
\subfigure{\includegraphics[width=0.07\textwidth, angle=0]{fLeg.pdf}}
\end{tabular}
\end{tabular}
\caption{(Color online) 
Fidelity $\overline{\cal F}$ as functions of $\Delta$ and $\phi$, for the same values of the parameters in Fig.~\ref{fig:Fid_InitCond}b and Fig.~\ref{fig:Fid_InitCond}d, that is: $g_1/\epsilon=g_2/\epsilon=0.2$ (a) and $g_1/\epsilon=g_2/\epsilon=0.5$ (b), while all the other parameters are $\Ket{\psi(0)}=\Ket{2}$, $\Omega/\epsilon=0.1$ and $\epsilon T=2\pi$.}
\label{fig:Fid_InitCond_Bar}
\end{figure*}

We have evaluated the fidelity ${\cal F}(T)$ and the normalized fidelity $\overline{{\cal F}}(T)$ in several conditions. In Fig.~\ref{fig:Fid_InitCond} we show the fidelity ${\cal F}(T)$ for different 
values of the coupling constants: $g_1/\epsilon=g_2/\epsilon=0.1$ (a), $g_1/\epsilon=g_2/\epsilon=0.2$ (b), $g_1/\epsilon=g_2/\epsilon=0.3$ (c), $g_1/\epsilon=g_2/\epsilon=0.5$ (d). In all plots we have $\omega/\epsilon=0.1$, $\epsilon T = 2\pi$ and the initial condition $\Ket{\psi(0)} = \Ket{2}$.
Fig.~\ref{fig:Fid_InitCond}a shows that ${\cal F} \approx 1$ almost everywhere, due to the low values of the coupling constants which make the perturbed evolution close to the unperturbed one.
From Fig.~\ref{fig:Fid_InitCond}b, \ref{fig:Fid_InitCond}c and \ref{fig:Fid_InitCond}d we get that, as expected, for large values of $\Delta$ (for example $\Delta/\epsilon>10$) an extended Zeno dynamics is predicted, irrespectively of $\phi$.  On the contrary, when $\Delta$ is moderately larger than $\epsilon$ ($1<\Delta/\epsilon<10$), a dependence of the fidelity from the phase $\phi$ is well visible from the figures. In particular, when $\phi$ is close to $\pi/2$, which means that the diagonal matrix element is essentially a decay rate, higher values of $\Delta$ are required to have a fidelity $\cal F$ close to unity. It is anyway interesting to investigate the reason for such a different behavior. To this end, it is useful to consider the normalized fidelity $\overline{{\cal F}}$ (see Fig.~\ref{fig:Fid_InitCond_Bar}) which on the one hand, reaches higher values for lower values of $\Delta$ and, on the other hand, allows to reveal a good agreement between the complete and the unperturbed dynamics even when $\phi\approx \pi/2$. 
In other words, there is essentially a good agreement between the unperturbed and the perturbed dynamics, the only difference being a general loss of probability due to the presence of dissipation. Therefore, up to a wave function renormalization, the two dynamics essentially coincide. This is in perfect agreement with our theoretical analysis. Indeed, for very large complex energy gaps we can use the first order perturbation treatment, which predicts a dynamics in the subspace $B$ which is very close to the one obtained in the absence of any interaction with $A$. When the complex energy gap is only moderately large, it is better to use the second order corrections. Since the eigenvalues associated to the subspace $B$ acquire imaginary parts, we predict a general decay for the projection of the wave function to the subspace $B$, which is the reason why the fidelity $\cal F$ lowers down. Nevertheless, up to such a complessive decay, the dynamics is essentially the one induced by $\mathbf{B}$, leading to higher values for the renormalized fidelity $\overline{\cal F}$.

\begin{figure*}[t]
\centering
\begin{tabular}{lr}
\begin{tabular}{cccc}
\subfigure[]{\includegraphics[width=0.30\textwidth, angle=0]{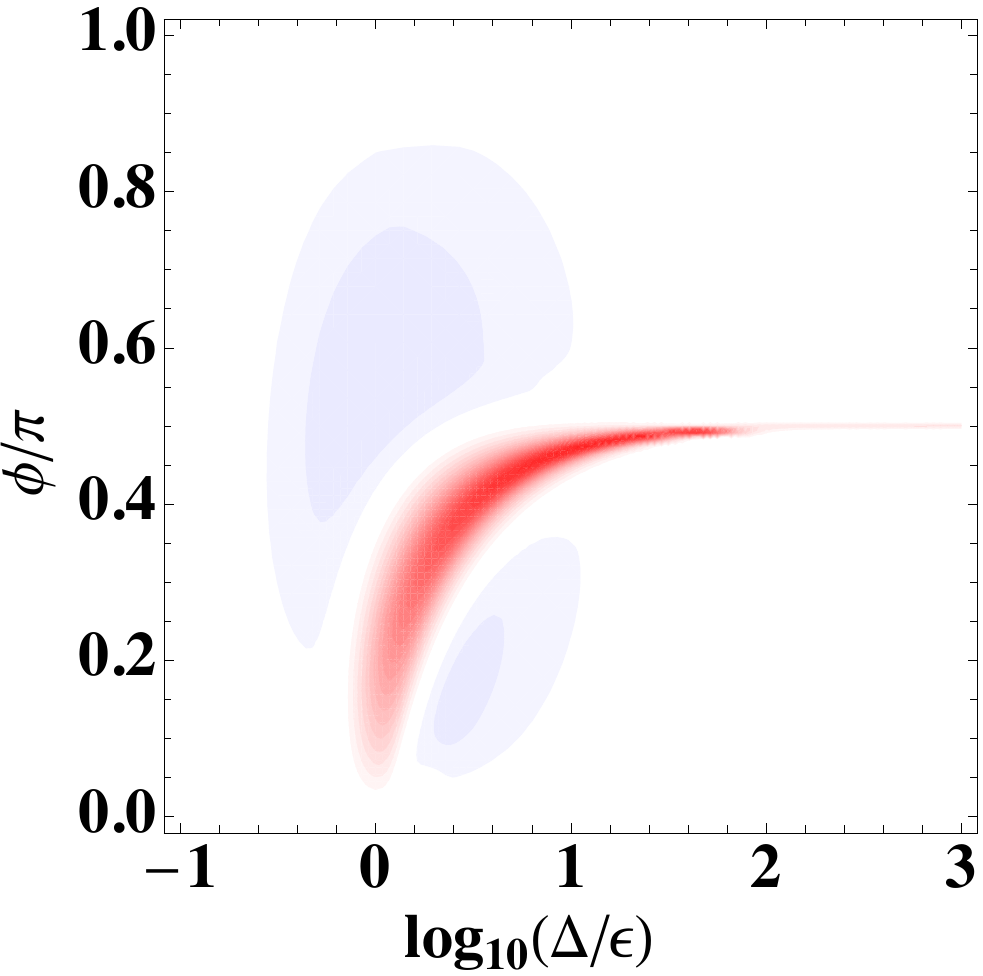}} &
\subfigure[]{\includegraphics[width=0.30\textwidth, angle=0]{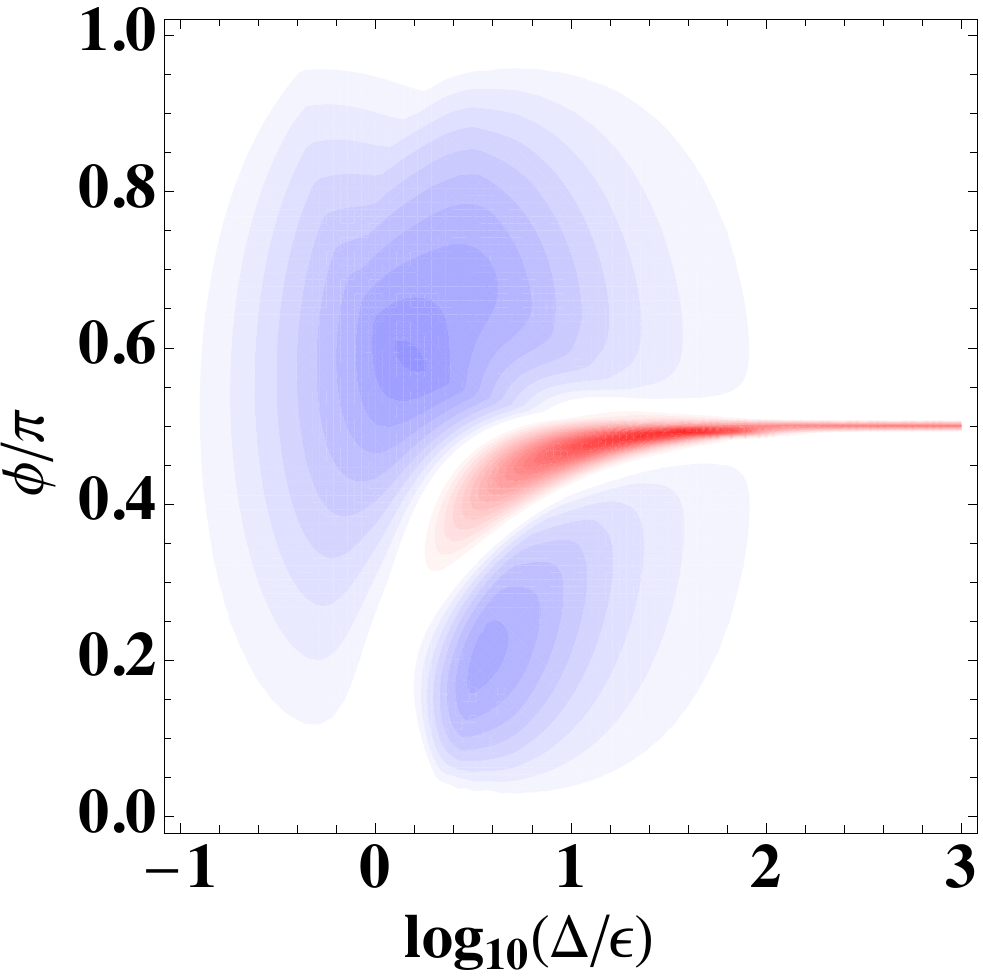}} \\
\end{tabular} & 
\begin{tabular}{c}
\subfigure{\includegraphics[width=0.08\textwidth, angle=0]{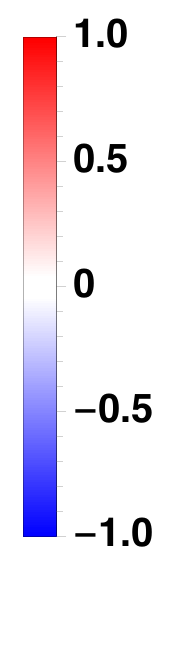}}
\end{tabular}
\end{tabular}
\caption{(Color online) 
Fidelity $\tilde{\cal F}$ as functions of $\Delta$ and $\phi$, for the same values of the parameters in Fig.~\ref{fig:Fid_InitCond}b and Fig.~\ref{fig:Fid_InitCond}d, that is: $g_1/\epsilon=g_2/\epsilon=0.2$ (a) and $g_1/\epsilon=g_2/\epsilon=0.5$ (b), while all the other parameters are $\Ket{\psi(0)}=\Ket{2}$, $\Omega/\epsilon=0.1$ and $\epsilon T=2\pi$.
}
\label{fig:Fid_Tilde}
\end{figure*}

In Fig.~\ref{fig:Fid_Tilde} we plot the fidelity $\tilde{\cal F}(T)$ in connection to the same values of the parameters used for Fig.~\ref{fig:Fid_InitCond}b and Fig.~\ref{fig:Fid_InitCond}d. It is well visible that in some zones the role of the dissipation is important for the dynamical confinement to be established (red zones), while in some other cases it can be (slightly) detrimental (light-blue zones). In extended areas its presence is more or less irrelevant (white zones). As it is expected, for $\phi=\pi/2$ we always have a red line in the region from moderately high to high values of $\Delta$, which corresponds to the occurrence of a pure Zeno effect. Moreover, a red zone is present also far from $\phi=\pi/2$, for moderately high values of $\Delta$. It is worth noting that, by plotting the function ${\cal F}(T) \times {\mathcal H}(\tilde{\cal F}(T))$, with ${\mathcal H}$ the Heaviside function, we have checked that in the biggest part of the red zones (where the decay helps the confinement) the fidelity ${\cal F}(T)$ is appreciably close to unity, i.e., the confinement occurs.

All plots previously shown, have been realized considering the initial condition $\Ket{\psi(0)}=\Ket{2}$. Nevertheless, we have checked that the figures coming from all the other initial conditions belonging to the subspace $\{\Ket{2}, \Ket{3}\}$ are very similar to those obtained starting from $\Ket{2}$. There are anyway some exceptions. Indeed, depending on the values of the parameters, some interaction-free states can be identified~\cite{ref:Chrusc2014,ref:Chrusc2015,ref:Chrusc2016}, for example states which are simultaneous eiegnstates of $H_0$ and $H_I$, so that their unperturbed evolution (which is a non evolution) coincided with the perturbed one (still no evolution). The analysis of this phenomenon is beyond the scope of the present work, but it deserves to be noted that such behavior must not be confused with the occurrence of an extended Zeno dynamics.

\section{Conclusions}\label{sec:conclusions}\label{sec:Conclusions}

In this paper we have analyzed the Zeno effect and, more specifically, the extended Zeno dynamics that can occur when a non-Hermitian Hamiltonian is considered. The model we have considered can be physically justified when the system is interacting with a zero-temperature reservoir and consists of a set of states decaying toward a group of lower (in terms of energy) ones, provided no coherent interaction between the lower and the upper states is present.
Studies including effective decays as imaginary parts of some diagonal entries of the Hamiltonian have been already presented, to show how continuous measurements, meant as decay processes, can allow for Zeno phenomena to occur as well as repeated pulsed measurement do.
When some diagonal matrix elements of the Hamiltonian are very large, a partitioning of the Hilbert space is realized, leading to the occurrence of a confined dynamics. This happens whether they are real or imaginary numbers. Nevertheless, we have presented a study of how the extended Zeno regime is reached when different phases are assumed for the very large diagonal entries. In particular, we have shown that when the extended Zeno dynamics is due to a decay (pure Zeno dynamics), its occurrence requires higher values than in the case where the Hilbert space partitioning is due to very large differences of proper energies. Our theoretical analysis, based on the perturbation treatment for non-Hermitian Hamiltonians, is well supported by numerical calculations of some appropriate parameters we have introduced, $\cal F$, $\overline{\cal F}$ and $\tilde{\cal F}$. In particular, the last one allows for identifying the parameter zones where the dissipation positively contributes to the confinement of the dynamics, that is, a proper Zeno effect occurs.

\appendix

\section{Back to the Master Equation}\label{sec:appME}

In this appendix we analyze again the derivation of the non-Hermitian Hamiltonian from the master equation in \eqref{eq:NStateSystemME}. To render the mathematical treatment less cumbersome we will consider the case where the jump operators have the simple form $\hat{X}_{kj} = \Ket{j}\Bra{k}$, thus connecting single decaying states $\Ket{k}$ with single ground states $\Ket{j}$ (it is not excluded that different decaying states are connected to the same ground state). The master equation then assumes the form
\begin{eqnarray}
\nonumber
\dot\rho &=& -\ii [H_S, \rho] \\
\nonumber
&+& \sum_{k=1}^{N}\sum_{j=N+1}^{N+Q} \gamma_{kj} (\Ket{j}\Bra{k} \rho \Ket{k}\Bra{j} - 1/2 \{\Ket{k}\Bra{k}, \rho\}) \,, \\
\end{eqnarray}
where $Q=\mathrm{tr}\hat{\Pi}_G$ is the dimension of the subspaces spanned by the ground states and $\Ket{j}$ with $j=N+1, ..., N+Q$ are such ground states the other states decay toward. 

The equations for the matrix elements of $\rho$ related to the states of the subspace $R$ ($\rho_{kk'}$ with $k,k'=1,..., N$) are:
\begin{eqnarray}
\nonumber
\dot\rho_{kk'} &=&   - \sum_{j=N+1}^{N+Q} \frac{\gamma_{kj} + \gamma_{k'j}}{2} \rho_{kk'}   \\
\nonumber
&-& \ii \sum_{l=1}^{N} \Bra{k} H_S \Ket{l} \rho_{lk'}  + \ii \sum_{l=1}^{N} \rho_{kl} \Bra{l} H_S \Ket{k'}  \,,\\
&& \qquad \qquad k, k' = 1,..., N\,.
\end{eqnarray}

It is well visible that the set of such equations is self-consistent in the sense that it does not involve the remaining matrix elements with one or two indices higher than $N$. Moreover, the equations are those associated to the non-Hermitian Hamiltonian $H$.
It is worth commenting that the set of equations for the matrix elements related to the ground states are not self-consistent, since such states receive population for the upper ones. Consider for example the following equations for the diagonal matrix elements:
\begin{eqnarray}
\nonumber
\dot\rho_{jj} &=&  \sum_{k=1}^{N} \gamma_{kj} \rho_{kk} \,,  \qquad j=N+1,..., N+Q\,.
\end{eqnarray}
It is clear that restricting our analysis to the ground subspace $G$ and describing the relevant dynamics through a non-Hermitian Hamiltonian would be impossible.

\section{Perturbation Treatment for Non-Hermitian Hamiltonians}\label{app:perturbation}

In this appendix we consider the perturbation theory for a non-Hermitian Hamiltonian which possesses non-degenerate subspaces.
When no degenerations are present, the left and right eigenvector problems can be directly solved in the usual way, with the only peculiarity that the left eigenvectors are not the adjoint of the corresponding right eigenvectors and must be found independently.
Let us assume $H = H_0 + \lambda H_1$ (both $H_0$ and $H_1$ can be non-Hermitian) and that we know the left and right eigenvectors of $H_0$: 
\begin{subequations}
\begin{eqnarray}
\Bra{u^{(0)}_k} H_0 = \epsilon^{(0)}_k \Bra{u^{(0)}_k}\,, \\ 
H_0\Ket{v^{(0)}_k} = \epsilon^{(0)}_k \Ket{v^{(0)}_k}\,,
\end{eqnarray}
with the bi-orthogonality condition
\begin{eqnarray}
\BraKet{u^{(0)}_k}{v^{(0)}_j} = \delta_{kj} \,.
\end{eqnarray}
\end{subequations}

The eigenvector equations for the total Hamiltonian read,
\begin{eqnarray}
(H_0 + \lambda H_1) \sum_n \lambda^n \Ket{v^{(n)}_k} = \sum_p \lambda^p \epsilon^{(p)}_k \sum_q \lambda^q \Ket{v^{(q)}_k}\,, \quad \\
(H_0 + \lambda H_1) \sum_n \lambda^n \Bra{u^{(n)}_k} = \sum_p \lambda^p \epsilon^{(p)}_k \sum_q \lambda^q \Bra{u^{(q)}_k}\,. \quad
\end{eqnarray}

By projecting the first of such equations through application of $\Bra{u^{(m)}_j}$ and the second equation through $\Ket{v^{(m)}_j}$, and equating the terms with the same order in $\lambda$, we straightforwardly obtain the equations for the corrections. In the following we give the correction up to second-order, which resembles the standard one (i.e., the one obtained for Hermitian operators) except for the fact that the ordinary \lq bra\rq\, are replaced by the left eigenvectors ($\Bra{v_k^{(n)}} \rightarrow \Bra{u_k^{(n)}}$):
\begin{subequations}
\begin{eqnarray}
\nonumber
\epsilon_k &=& \epsilon_k^{(0)} + \lambda \Bra{u^{(0)}_j} H_1 \Ket{v^{(0)}_j}  \\
\nonumber
&+& \lambda^2 \sum_{j\not=k} \frac{\Bra{u^{(0)}_j} H_I \Ket{v_k^{(0)}} \Bra{u^{(0)}_k} H_I \Ket{v_j^{(0)}} }{\epsilon_k^{(0)} - \epsilon_j^{(0)}}  \\
&+& o(\lambda^3) \,,
\end{eqnarray}

\begin{eqnarray}
\nonumber
&&\!\!\!\!\!\!\!\!\! \Ket{v_k} = \Ket{v^{(0)}_k} + \lambda \sum_{j\not=k} \frac{\Bra{u^{(0)}_j} H_I \Ket{v_k^{(0)}}}{\epsilon_k^{(0)} - \epsilon_j^{(0)}} \Ket{v_j^{(0)}}  \\
\nonumber
&+& \lambda^2   \sum_{m, n \not=k}  \frac{  \Bra{u^{(0)}_n} H_I \Ket{v_m^{(0)}}  \Bra{u^{(0)}_m} H_I \Ket{v_k^{(0)}}  }{ (\epsilon_k^{(0)} - \epsilon_n^{(0)}) (\epsilon_k^{(0)} - \epsilon_m^{(0)}) } \Ket{v_m^{(0)}}     \\ 
\nonumber
&-& \lambda^2   \sum_{n \not=k}  \frac{  \Bra{u^{(0)}_k} H_I \Ket{v_k^{(0)}}  \Bra{u^{(0)}_n} H_I \Ket{v_k^{(0)}}  }{ (\epsilon_k^{(0)} - \epsilon_n^{(0)})^2 } \Ket{v_n^{(0)}}     \\ 
\nonumber
&-& \frac{\lambda^2}{2}   \sum_{n \not=k}  \frac{  \Bra{u^{(0)}_k} H_I \Ket{v_n^{(0)}}  \Bra{u^{(0)}_n} H_I \Ket{v_k^{(0)}}  }{ (\epsilon_k^{(0)} - \epsilon_n^{(0)})^2} \Ket{v_k^{(0)}}    \\
&+& o(\lambda^3)  \,, 
\end{eqnarray}\\

\begin{eqnarray}
\nonumber
&&\!\!\!\!\!\!\!\!\! \Bra{u_k} = \Bra{u^{(0)}_k} + \lambda \sum_{j\not=k} \frac{\Bra{u^{(0)}_k} H_I \Ket{v_j^{(0)}}}{\epsilon_k^{(0)} - \epsilon_j^{(0)}} \Bra{u_j^{(0)}}  \\
\nonumber
&+& \lambda^2   \sum_{m, n \not=k}  \frac{  \Bra{u^{(0)}_n} H_I \Ket{v_m^{(0)}}  \Bra{u^{(0)}_m} H_I \Ket{v_k^{(0)}}  }{ (\epsilon_k^{(0)} - \epsilon_n^{(0)}) (\epsilon_k^{(0)} - \epsilon_m^{(0)}) } \Bra{u_m^{(0)}}     \\ 
\nonumber
&-& \lambda^2   \sum_{n \not=k}  \frac{  \Bra{u^{(0)}_k} H_I \Ket{v_k^{(0)}}  \Bra{u^{(0)}_n} H_I \Ket{v_k^{(0)}}  }{ (\epsilon_k^{(0)} - \epsilon_n^{(0)})^2 } \Bra{u_n^{(0)}}     \\ 
\nonumber
&-& \frac{\lambda^2}{2}   \sum_{n \not=k}  \frac{  \Bra{u^{(0)}_k} H_I \Ket{v_n^{(0)}}  \Bra{u^{(0)}_n} H_I \Ket{v_k^{(0)}}  }{ (\epsilon_k^{(0)} - \epsilon_n^{(0)})^2} \Bra{u_k^{(0)}}     \\
&+& o(\lambda^3)  \,. 
\end{eqnarray}
\end{subequations}

\end{document}